# The Interstellar H Flow: Updated Analysis of SOHO/SWAN Data

Rosine Lallement*, Eric Quémerais*, Dimitra Koutroumpa$ , Jean-Loup Bertaux*, Stéphane Ferron*, Walter Schmidt§ and Philippe Lamy#

*UVSQ, LATMOS/IPS, CNRS/UMR 8190, BP3, 91371 Verrières-le-Buisson, France
$ Code 662, NASA/Goddard Space Flight Center, Greeblet, MD 20771, USA
$ Finnish Meteorological Institute, Helsinki, Finland
#Université de Provence LAM, CNRS/UMR 6110, Technopôle de Marseille-Etoile, France

**Abstract.** We update two kinds of results obtained with the SWAN instrument on board SOHO. First, we use H cell data recorded in 2001 and derive the H flow direction in the same way the study was done at solar minimum. We again compare with the Helium flow direction and doing so correct for the coordinate system change between the Ulysses and SOHO missions. The deflection plane we obtain is compatible with our previous result within error bars, confirming the potential predominant role of the interstellar magnetic field. In a second part, we extend the computation of the interstellar H ionization as a function of heliographic latitude and time, a quantity which reflects closely the SW flux latitudinal structure. The pattern for the present solar minimum is strikingly different from the previous minimum, with a much wider slow solar wind equatorial belt which persists until at least 2008. Comparing with synoptic LASCO/C2 electron densities we infer from a preliminary study that the acceleration of the high speed solar wind occurs at a higher altitude during this minimum compared to the previous one, a difference expansion models must be able to reproduce.

**PACS:** Replace this text with PACS numbers; choose from this list: http://www.aip.org/pacs/index.html

## INTRODUCTION

The SWAN instrument on board SOHO is recording full-sky maps of the diffuse Ly-alpha background emission with a resolution of 1x1 deg$^2$, as well as the fraction of this emission that is blocked out by an absorption cell when this cell is filled with atomic hydrogen. The cell is alternatively activated and turned OFF in such a way the ratio $I_{ON}/I_{OFF}$ (attenuated over un-attenuated brightness) is measured at almost the same time in each direction.

Except for a contamination by the bright early-type stars, most of the signal is the solar Lyman-alpha radiation backscattered by interstellar H atoms which are permanently flowing within the heliosphere. The relative motion of the Sun with respect to the surrounding interstellar gas (about 25 km/s) is of the same order than the Earth orbital speed, which fortunately modulates strongly the relative motion between the H cell (or the SOHO spacecraft) and the emitting interstellar gas. In other words, according to the viewing direction and to the date the cell produces an absorption line at different locations of the Ly-alpha emission line and a good scanning of this line can be performed (for details about the instrument and the observing strategies see e.g. [1,2,3]).

The distribution of the H atoms is deeply affected by the solar wind through charge transfer with solar ions and also photo-ionization, which allows to infer solar wind fluxes as a function time and heliolatitude. Here we update the solar wind large scale distribution derived from SWAN brightness maps (§3). In parallel, the H cell acts as a negative spectrometer and allows a precise determination of the flow dynamics. In the next § we update the determination of the H flow direction inferred from the cell absorption maps.

## H CELL DATA: THE H FLOW DEFLECTION

The H flow is made of pristine interstellar atoms, i.e. atoms having reached the inner heliosphere without any charge transfer collision, and of the so-called secondary atoms, former interstellar protons having been neutralized during charge-transfer with neutrals. Since interstellar protons are following trajectories which lead them to decelerate during their approach of the heliopause and get deviated to flow around the heliosphere, the newly created H atoms have dynamical properties which keep the imprints of these interaction processes with the heliospheric boundary. Indeed, the average H flow has been found to be slower (by about 4 km/s) than the interstellar helium flow, in agreement with the lack of perturbation of the helium atoms which have a much smaller charge-exchange cross-section with protons and helium ions. The velocity dispersion of H (an average temperature of 13,000K [3]) also reflects the existence of the two flows, while interstellar helium keeps its initial external temperature of about 6,000K (as it is measured towards the nearby stars). However, the H flow has also been found to be deviated by a few degrees with respect to He, a fact interpreted as the signature of an asymmetry of the outer heliosphere under the effect of an inclined interstellar magnetic field [4].

This deflection was inferred from the analysis of data recorded at the beginning of the mission, in 1996, i.e. at solar minimum. Since there are large solar wind anisotropies at low activity that could induce departures from axial symmetry of the heliosphere (see also the possible influence of the IP field [5]), it is useful to reproduce the same analysis at solar maximum to help to disentangle interstellar from solar magnetic field effects.

Details of the analysis can be found in [4]. In brief, we use a forward model of a mono-kinetic flow and adjust the model parameters to fit the locations of the maximum absorption by the cell (the Zero Doppler Shift directions), i.e. where the cell is absorbing right at the center of the emission line. We have used here 2001 data (16 maps) and varied the flow speed, radiation pressure, and flow ecliptic coordinates (half-degree grid). Fig. 1 shows the resulting $\chi^2$ iso-contours and the location of the best-fit flow direction together with the previous results of the solar minimum study. The new direction is slightly different from the direction found with the same method for 1996 data, with a 0.4° longitude increase. This is a very small change, which shows that the deflection of the H flow is not significantly different in 2001 compared to 1996, and that it is governed by processes mostly independent from the cycle phase. In the previous study we had also made use of an additional, completely different and model-independent method based on line profile deconvolution [2], which led to a slightly different direction, also shown in Fig 1, differing by about 0.5°. We had estimated the uncertainty on the direction to be of the order of the difference between the two independent determinations. Since we have used the first method only for 2001, we now have 3 determinations, from which we estimate the arrival direction to be $(\lambda,\beta)_H$= (252.5± 0.7 °, +8.9±0.5 °). Using this result we can re-estimate the H deflection plane (HDP), by comparison with the He flow [6]. We use here the J2000 coordinate system for both SOHO and Ulysses, correcting for an error in our previous study due to a misunderstanding between the Ulysses team and ourselves, and noticed by [7]. Using $(\lambda,\beta)_{He}$= (255.4± 0.5 °, +5.2 ±0.2 °), we find that the HDP projection is inclined by 38±7° from the North Ecliptic Pole direction, as measured clockwise looking towards the heliosphere nose, which is compatible within uncertainties with the angle derived from the source distribution of the 2-3 KHz emission [8] under the assumption that the emissions are due to the interaction between propagating IP shocks and the heliopause, and emissions are localized where the IS field is nearly perpendicular to the shock normal.

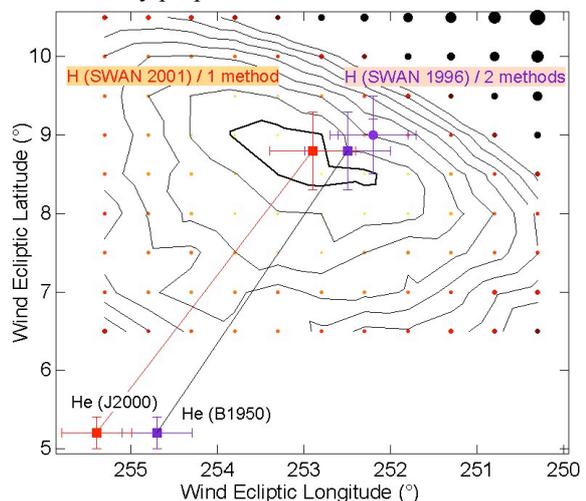

**FIGURE 1.** Determination of the H flow direction: chi-square iso-contours for the data/model adjustment in 2001. Dots represent the computation grid. The best-fit direction found for the 1996 data from the two different methods are shown for comparison, as well as the Helium flow direction.

## BRIGHTNESS MAPS: SOLAR WIND LATITUDINAL STRUCTURE

The H distribution and therefore the sky Ly-α brightness pattern reflects the ionization of the neutral gas by the solar wind (the main effect) and the radiation. A forward model allowing for a latitude-dependent ionization has been adjusted to the SWAN maps, and we present here a preliminary update of this modeling to cover up the latest measurements. For details about the model and the method see [9]. Fig. 2 displays the results from 830 maps between 1996 and 2009. The quantity shown is the H global ionization rate τ(H) at 1AU, normalized here to the solar equator sector. Note that the polar rate is poorly defined due to the short time atoms spend traveling above the solar pole.

The ionization rate is the sum of the photo-ionization (about 20%) and the ionization by charge-transfer with the solar wind ions (mostly protons). The latter term is the product of the flux by the cross-section σ, which has a measured dependence on the velocity.

$$\tau_{total}(H) = \tau_{phot}(H) + \tau_{SW}(H)$$
$$\tau_{SW}(H) = \phi_{SW}\sigma(V_{SW})$$

Using proxys for the EUV flux and IPS data for the mean velocity, the global rate can be converted into solar wind fluxes. The pattern for the flux (not shown here) is very similar to the global rate pattern shown here. While in 1996-97 the slow solar wind low-latitude belt is 25° wide only, the rest being occupied by low flux high speed wind, the situation in 2006-2009 is very different, with a much wider belt (up to 90-100°). This is in agreement with Ulysses electron and ion data and IPS speed latitudinal structure [10,11,12,13], but here we additionally show that this is a permanent, persisting feature which is also reflected in the global fluxes themselves.

There is a strong similarity between the H ionization pattern and the 6 $R_s$ LASCO-C2 coronal electron densities, displayed in Figure 3. High densities correspond to high ionization by the low speed wind, while low densities are found in the polar holes, sources of the fast wind. The combination of the SWAN SW fluxes deduced from the ionization rates and the LASCO data between 2.5 and 7 $R_s$ has led to a very interesting result [14] for the 1996 minimum: the fast wind has almost reached its final speed at 6 Rs, while at the same altitude the slow wind has only reached about half of its final speed. This had been inferred in some particular cases, but the SWAN/LASCO combination demonstrates that it is a general property valid at any time and at all latitudes. Such a study, this time for the present minimum is in progress, but it is possible to foresee a peculiar result, by simply using mass conservation along flow tubes. As a matter of fact, the LASCO maps show that the electron density above 40° latitude is higher during this minimum than in 96 (note also that there is a mid-latitude minimum, not present in 96). On the other hand, it is clear from a number of data that the high-latitude SW fluxes are 20-25% below the previous minimum fluxes [10,11,12]. This is confirmed by SWAN, using the normalized values of Figure 2 and in-ecliptic SW data. The terminal speeds finally are similar, or only a few % less, as shown by SW data and IPS. If the coronal velocity profiles were unchanged, lower fluxes and identical final speeds would imply lower densities at every altitude, which is contradicted by LASCO. On the contrary, LASCO data imply that the velocity profile at high latitude is different from the last minimum, with less acceleration (leading to higher densities) at 6-7 Rs. That the acceleration of the fast wind occurs at a higher altitude (above 6 Rs) in polar holes, in conjunction with lower coronal magnetic fields, SW fluxes and temperatures is an interesting test for expansion models .

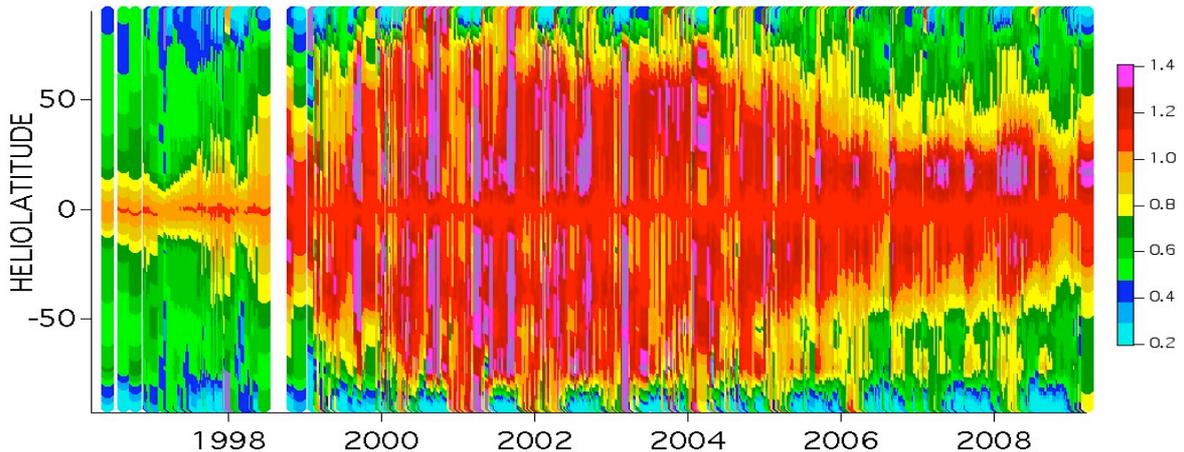

**FIGURE 2.** Interstellar H ionization rate from 830 SWAN intensity maps. Most of the ionization is due to charge transfer with solar wind ions. High speed (low flux) winds from coronal holes appear in blue and green, low speed (high flux) winds in red. The rate here is normalized to the equatorial value. The present minimum is characterized by a very broad slow wind equatorial belt, which is shown to be a permanent feature persisting until 2008.

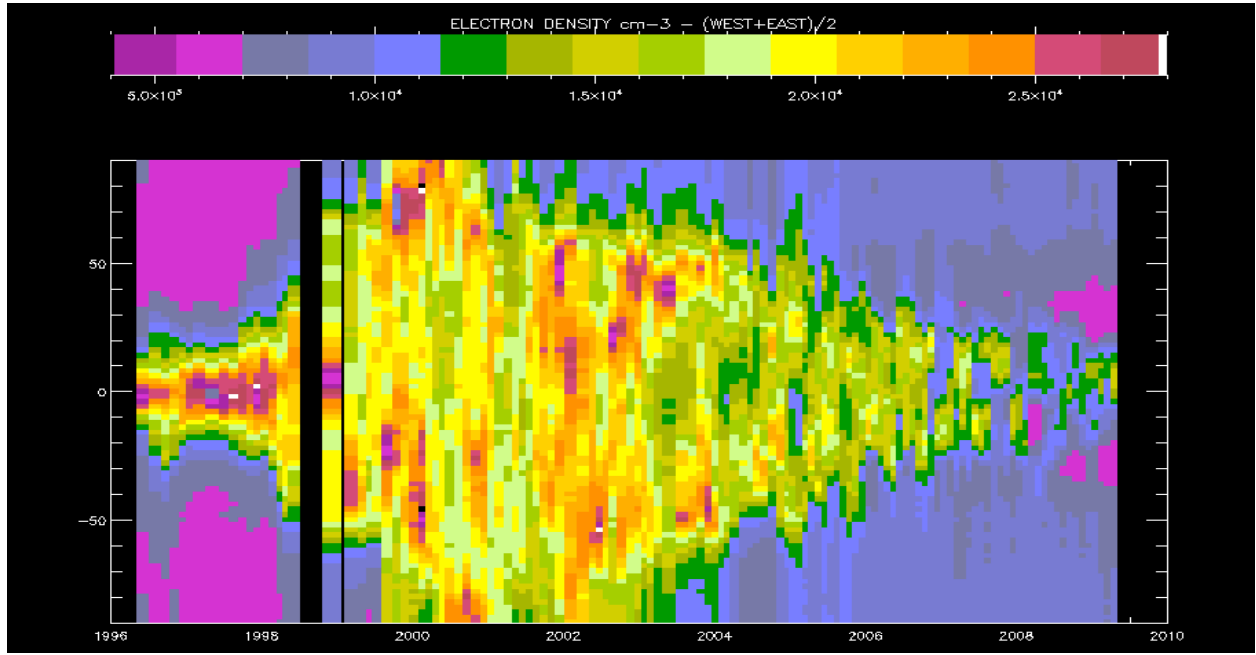

**FIGURE 3**: SOHO LASCO/C2 electron densities at 6 solar radii. There is a strong similarity between the SWAN ionization rates (Fig. 2) and the coronal densities, with a broader high density equatorial band during the present minimum. Note however the peculiar mid-latitude minima in 2007-2009, absent in 96-98.

## ACKNOWLEDGMENTS

SOHO is a mission of international cooperation between ESA and NASA. SWAN was financed in France by CNES with support from CNRS and in Finland by TEKES and the Finnish Meteorological Institute.
We thank the organisers of this productive and enjoyable meeting !